\documentclass[usenatbib]{mn2e}

\usepackage{graphicx}
\usepackage{amsmath}
\usepackage{lscape}
\usepackage{afterpage}
\usepackage{soul}
\usepackage{color}
\usepackage{url}
\newcommand{\angstrom}{\textup{\AA}}

\title[The HLX candidate in IC 4320]{The hyperluminous X-ray source
candidate in IC 4320: another HLX bites the dust}

\author[A. D. Sutton et al.]{Andrew D. Sutton,$^{1,2}$\thanks{Email:
andrew.d.sutton@nasa.gov} Timothy P. Roberts,$^1$ Jeanette C.
Gladstone,$^3$\and Dominic J. Walton$^4$\\
 \\
$^1$Department of Physics, University of Durham, South Road, Durham, DH1
3LE, UK\\
$^2$Astrophysics Office, NASA Marshall Space Flight Center, ZP12, Huntsville, Al 35812, USA\\
$^3$Department of Physics, University of Alberta, 11322-89 Avenue,
Edmonton, Alberta T6G 2G7, Canada\\
$^4$Space Radiation Laboratory, California Institute of Technology,
Pasadena, CA 91125, USA}
\pagerange{\pageref{firstpage}--\pageref{lastpage}} \pubyear{2014}

\def\la{\mathrel{\hbox{\rlap{\hbox{\lower4pt\hbox{$\sim$}}}{\raise2pt\
hbox{$<$}}
}}}
\def\ga{\mathrel{\hbox{\rlap{\hbox{\lower4pt\hbox{$\sim$}}}{\raise2pt\
hbox{$>$}}
}}}

\begin{document}

\maketitle

\label{firstpage}

\begin{abstract}
The known members of the class of hyperluminous X-ray sources (HLXs) are
few in number, yet they are of great interest as they are regarded as
the likeliest intermediate-mass black hole (IMBH) candidates amongst the
wider population of ultraluminous X-ray sources (ULXs).  Here we report
optical photometry and spectroscopy of a HLX candidate associated with
the galaxy IC 4320, that reveal it is a background AGN.  We discuss the
implications of the exclusion of this object from the small number of
well-studied HLXs, that appears to accentuate the difference in
characteristics between the good IMBH candidate ESO 243-49 HLX-1 and the
small handful of other HLXs.
\end{abstract}


\section{Introduction}\label{intro}

Hyperluminous X-ray sources (HLXs) are a subset of the brightest
ultraluminous X-ray sources (ULXs).  They are defined as point sources
with X-ray luminosities in excess of $10^{41}~{\rm erg~s^{-1}}$,
located away from the nucleus of external galaxies
\citep{gao_etal_2003}.  However, they remain poorly understood
compared to their lower-luminosity relations, for which substantial
progress has been made in recent years.  In particular, it has now
been demonstated that at least three individual ULXs are powered by
super-Eddington accretion onto stellar-mass black holes, with these
objects located in the galaxies M31, M101 and NGC 7793
(\citealt{middleton_etal_2013, liu_etal_2013, motch_etal_2014}).  The
last of these detections is crucial in understanding the wider
population of ULXs as this object shows that the key spectral and
behavioural differences between ULXs and sub-Eddington stellar-mass
black holes (e.g. \citealt{gladstone_etal_2009, sutton_etal_2013b};
see also \citealt{feng_and_soria_2011} and references therein, with
recent confirmation of the spectral differences coming from {\it
NuSTAR} observations, e.g.
\citealt{bachetti_etal_2013,walton_etal_2013_circinus}) manifest at
super-Eddington rates.  However, it is not yet clear whether HLXs can
be produced in this way.  It is therefore intriguing that luminosities
up to $\sim 2 \times 10^{41}~{\rm erg~s^{-1}}$ may be possible, from a
combination of the most massive stellar remnant black holes (up to
$\sim 80 ~M_{\odot}$ in regions of low metallicity;
\citealt{zampieri_and_roberts_2009, belczynski_etal_2010,
mapelli_etal_2010}), and maximal super-Eddington accretion rates
($\sim 20 ~L_{\rm Edd}$ from a face-on disc;
\citealt{ohsuga_and_mineshige_2011}).

Alternatively, HLXs may harbour a new class of black hole -- the
intermediate-mass black holes (IMBHs;
\citealt{colbert_and_mushotzky_1999}; $10^2$--$10^4~M_{\odot}$).  To
produce the observed X-ray luminosities while accreting in standard
sub-Eddington states requires HLXs to host IMBHs with masses of $> 10^3
~M_{\odot}$.  Indeed, there is arguably good evidence that this is the
case in the most luminous HLX, ESO 243-49 HLX-1 ($L_{\rm X, peak} \sim
10^{42}~{\rm erg~s^{-1}}$, \citealt{farrell_etal_2009}; although see
\citealt{king_and_lasota_2014} and \citealt{lasota_etal_2015} for an 
alternative scenario).  This
source exhibits regular outbursts, which appear to have a fast rise,
exponential decay (FRED) profile, repeated on a timescale of $\sim
1~{\rm year}$ (\citealt{lasota_etal_2011}; but see
\citealt{godet_etal_2013}).  Most crucially, its behaviour on the
hardness-intensity diagram as it progresses through these outbursts
appears to mimic sub-Eddington black hole binaries
\citep{servillat_etal_2011}.  It has been suggested that the putative
IMBH in ESO 243-49 HLX-1 may have originated in the nucleus of a stripped satellite
dwarf galaxy, whose host has been disrupted in an encounter with the
larger galaxy, or a large black hole recoiling from a close encounter
with the nuclear black hole of ESO 243-49 \citep{soria_etal_2013}.  

There are very few other sources that join ESO 243-49 HLX-1 in the
category of confirmed or even potential HLXs.  The best candidates include M82 X-1
\citep{matsumoto_etal_2001}, Cartwheel N10
\citep{gao_etal_2003,wolter_and_trinchieri_2004}, 2XMM
J011942.7$+$032421 in NGC 470 and 2XMM J134404.1$-$271410 in IC 4320
\citep{walton_etal_2011b,sutton_etal_2012}, the last of which we
consider further in this paper.  Several other candidate HLXs have been
mooted, but the evidence for these is less conclusive.  They include n40
in NGC 5775, which was estimated by \cite{ghosh_etal_2009} to have an
intrinsic X-ray luminosity of $L_{\rm X} \sim 10^{41}~{\rm erg~s^{-1}}$,
although its observed luminosity was nearer $7.5 \times 10^{40} ~\rm
erg~s^{-1}$.   Also included is CXO J122518.6$+$144545, which has an
intrinsic 0.5--10 keV luminosity of $L_{\rm X} \sim 2.7 \times
10^{41}~{\rm erg~s^{-1}}$, if it is at the distance of its presumed host
galaxy.  However, in this case a blue Type IIn supernova or a recoiling
SMBH offer viable alternative solutions \citep{jonker_etal_2010}. 
\cite{davis_and_mushotzky_2004}  identified 2XMM J072647.9$+$854550 in
NGC 2276 as a potential HLX, at $L_{\rm X} \sim 1.1 \times 10^{41}~{\rm
erg~s^{-1}}$.  But, this object has been both resolved into a triplet of
ULXs by {\it Chandra\/} observations, and its luminosity revised down
based on a more recent distance measurement, such that it is no longer
hyperluminous \citep{wolter_etal_2011, sutton_etal_2012}, although a 
recent X-ray/radio fundamental plane measurement highlights this ULX as a 
very strong IMBH candidate \citep{mezcua_etal_2015}.  Several other
HLX candidates have been categorically ruled out.  XMMU
J132218.3$-$164247 \citep{miniutti_etal_2006} turned out to be a type 1
QSO at $z \sim 1$ \citep{dadina_etal_2013}, and \cite{sutton_etal_2012}
identified two further HLX candidates from the 2XMM catalogue -- 2XMM
J120405.8$+$201345 and 2XMM J125939.8$+$275718 -- as background QSOs.

Given the scarcity of these objects, it is important that we properly
scrutinise each potential source.  Here we focus on 2XMM
J134404.1$-$271410, the HLX candidate in IC 4320.  At an X-ray
luminosity of $L_{\rm X} = 3.5^{+0.2}_{-0.3} \times 10^{41}~{\rm
erg~s^{-1}}$ (if it is at the distance of IC 4320), it is potentially
second only in peak luminosity amongst all HLXs to ESO 243-49 HLX-1. 
Even at this extreme luminosity, it has been observed in an apparent
low/hard X-ray state, making it an excellent IMBH candidate
\citep{sutton_etal_2012}.  As such, it is critical to confirm whether
this source is indeed associated with IC 4320.

In this work, we report on VLT imaging and spectroscopic observations of
2XMM J134404.1$-$271410.  We present the data in section~\ref{results},
and show that they rule this source out as a potential HLX.  In section
\ref{discussion} we discuss the physical interpretation of the dwindling
HLX population.

\section{Analysis and results}
\label{results}

\begin{table}
\centering
\caption{VLT VIMOS exposures}
\begin{tabular}{ccccc}
\hline
Obs. ID$^a$ & Filter$^b$ & Date$^c$ & $N_{\rm exp} ^d$ & $t_{\rm exp}
^e$ \\
\hline
822880 & {\it U} & 2013-02-12 09:09:00 & 3 &  69 \\
       & {\it I} & 2013-02-12 09:17:21 & 3 & 301 \\
786873 & {\it B} & 2013-02-14 08:09:36 & 3 & 245 \\
       & {\it V} & 2013-02-14 08:26:35 & 3 & 344 \\
786878 & {\it R} & 2013-02-15 08:20:12 & 3 & 501 \\
\hline
\end{tabular}
\label{vimos_obs}
\begin{minipage}{\linewidth}
Details of the VIMOS photometric exposures of 2XMM J134404.1$-$271410,
taken using VLT UT3, as part of VLT observing run 090.D-0300(A).
$^a$Observation identifier.
$^b$VLT VIMOS broad band filter in which the observation was taken.
$^c$Start date and time of the first exposure in the series (YYYY-MM-DD
HH:MM:SS).
$^d$Total number of consecutive exposures with the same filter.
$^e$Exposure time of each of the $N_{\rm exp}$ exposures (s).
\end{minipage}
\end{table}

\begin{figure}
\begin{center}
\includegraphics[width=8.5cm]{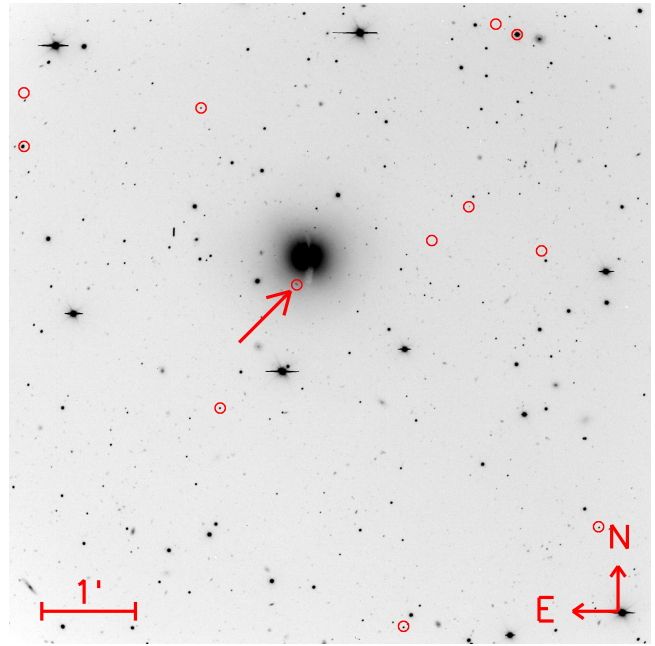}
\caption{Logarithmically scaled {\it V}-band VIMOS image of the region
around IC 4320.  The positions of nearby X-ray sources are shown with
arbitrarily-sized red circles.  These were identified from the {\it
Chandra} data, and used to correct the relative astrometry of the
images.  The red arrow indicates the {\it Chandra} position of 2XMM
J134404.1$-$271410.}
\label{astrometry}
\end{center}
\end{figure}

\begin{figure}
\begin{center}
\includegraphics[width=6.5cm]{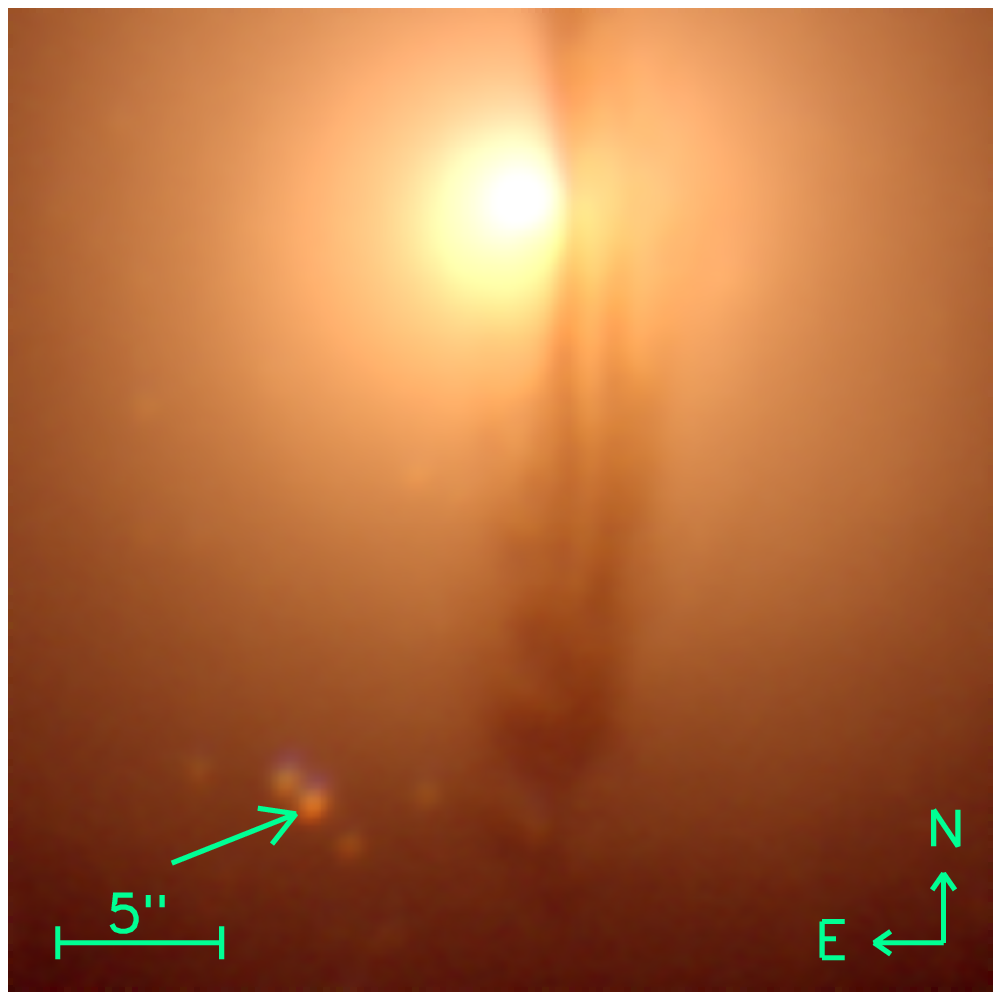}

\vspace{0.5cm}

\includegraphics[width=6.5cm]{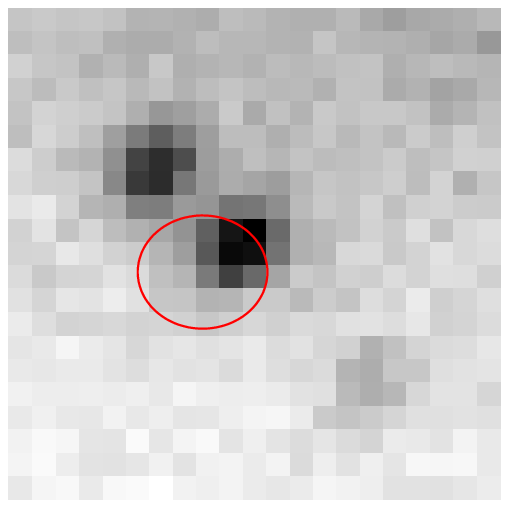}
\caption{({\it top}) Smoothed, true colour image showing the region around 2XMM
J134404.1$-$271410.  The RGB colours correspond to the {\it RVB} VLT
VIMOS bands, and the green arrow shows the counterpart, which was the
target of a subsequent series of VLT FORS2 long slit spectral
observations. ({\it bottom}) Close-up, unsmoothed VLT VIMOS V-band image of the region 
close to the ULX. The red ellipse indicates the $3 \sigma$ {\it Chandra} uncertainty 
region of the X-ray point source, after correcting for the relative astrometry of 
the X-ray and optical images.}
\label{image}
\end{center}
\end{figure}

\begin{table}
\centering
\caption{VLT FORS2 exposures}
\begin{tabular}{ccccc}
\hline
Obs. ID$^a$ & Date$^b$ & $N_{\rm exp} ^c$ & $t_{\rm exp} ^d$ \\
\hline
1000792 & 2014-03-10 06:15:15 & 2 & 1360 \\
1000795 & 2014-03-12 05:29:45 & 2 & 1360 \\
\hline
\end{tabular}
\label{fors2_obs}
\begin{minipage}{\linewidth}
Details of the FORS2 long slit spectral exposures of 2XMM
J134404.1$-$271410, taken using VLT UT1, as part of VLT observing run
092.D-0212(A).  All of the exposures were taken using the 0.7 arcsecond
slit, the GRIS300V+10 grism and the GG435 filter.
$^a$Observation identifier.
$^b$Start date and time of the first exposure in the series (YYYY-MM-DD
HH:MM:SS).
$^c$Total number of consecutive exposures.
$^d$Exposure time of each of the $N_{\rm exp}$ exposures (s).
\end{minipage}
\end{table}

The region around IC 4320 was observed with VLT VIMOS in imaging mode,
using the {\it UBVRI} filters.  Details of these observations are given
in Table \ref{vimos_obs}, and they are available in the ESO data
archive.\footnote{\url{archive.eso.org}}  The science exposures and
associated calibration files were reduced using the standard tools in
the VLT VIMOS pipeline (version 2.9.1), along with the common pipeline
library (CPL; version
6.1.1),\footnote{\url{www.eso.org/sci/software/pipelines/}} in {\sc
gasgano} (version
2.4.3).\footnote{\url{www.eso.org/sci/software/gasgano.html}}  In order
to identify any counterpart to the HLX candidate, we needed to align the
VIMOS image with an X-ray image.  To do this, we used data from a 50 ks
{\it Chandra} ACIS-S observation (observation ID 12989), previously
reported in \cite{sutton_etal_2012}.  We detected X-ray sources in a
0.5-7\,keV image using {\sc wavdetect} in {\sc ciao} (version
4.4)\footnote{\url{cxc.harvard.edu/ciao}}, focussing on a 6 arcminute
region around 2XMM J134404.1$-$271410.  Excluding the HLX candidate, 11
X-ray sources were detected in the same VIMOS quadrant as the target. 
These sources were used to align the relative astrometry of the X-ray
and optical images (Figure~\ref{astrometry}), via the {\sc iraf} tools
{\sc ccfind}, {\sc ccmap} and {\sc ccwcs}.  In this way, we were able to
constrain the relative right ascension and declination corrections to 
$\pm$ 0.181 and 0.158
arcsecond respectively.  These values were combined with the $\sim 0.01$
arcsecond error in the centroid position of the {\it Chandra} source to
calculate a combined ($3 \sigma$) uncertainty region with semi-major and
minor axes of $\sim 0.5$ arcsecond.  Having done this, we were able to
identify an optical source that was coincident with 2XMM
J134404.1$-$271410  (Figure~\ref{image}).  This counterpart had
(Galactic-reddening-corrected) apparent Vega magnitudes of $m_{U} = 24.5 \pm
0.6$, $m_{B} = 23.16 \pm 0.07$, $m_{V} = 22.67 \pm 0.05$, $m_{R} = 21.51
\pm 0.05$ and $m_{I} = 20.8 \pm 0.2$.  If the source is indeed
associated with IC 4320, then it would have a distance modulus of $\sim
34.89$, and the counterpart would be both brighter and redder than that
of ESO 243-49 HLX-1 \citep{soria_etal_2012,farrell_etal_2012}.

Having identified an optical counterpart, it was critical to confirm
whether it was indeed associated with IC 4320.  To this end, we were
awarded VLT FORS2 observing time to obtain a long slit spectrum, and
hence measure the redshift of the source.  Details of the FORS2
exposures are given in Table~\ref{fors2_obs}; these will be available in
the ESO archive following the 1 year proprietary period.  The
observations and associated calibration files were processed using {\sc
fors\_bias}, {\sc fors\_calib} and {\sc fors\_science}, as part of the
FORS pipeline (version 4.11.13), with the CPL (version 6.3). Initially, 
{\sc fors\_science} was used to process standard star observations of 
Feige 66 to produce spectral response curves for the instrument configuration. 
These spectral response curves were then used as inputs to the {\sc fors\_science} 
pipeline when processing the science exposures, to allow us to produce 
flux-calibrated spectra for each exposure. Finally, we median combined the
individual flux-calibrated science spectra.

\begin{figure*}
\begin{center}
\includegraphics[width=16cm]{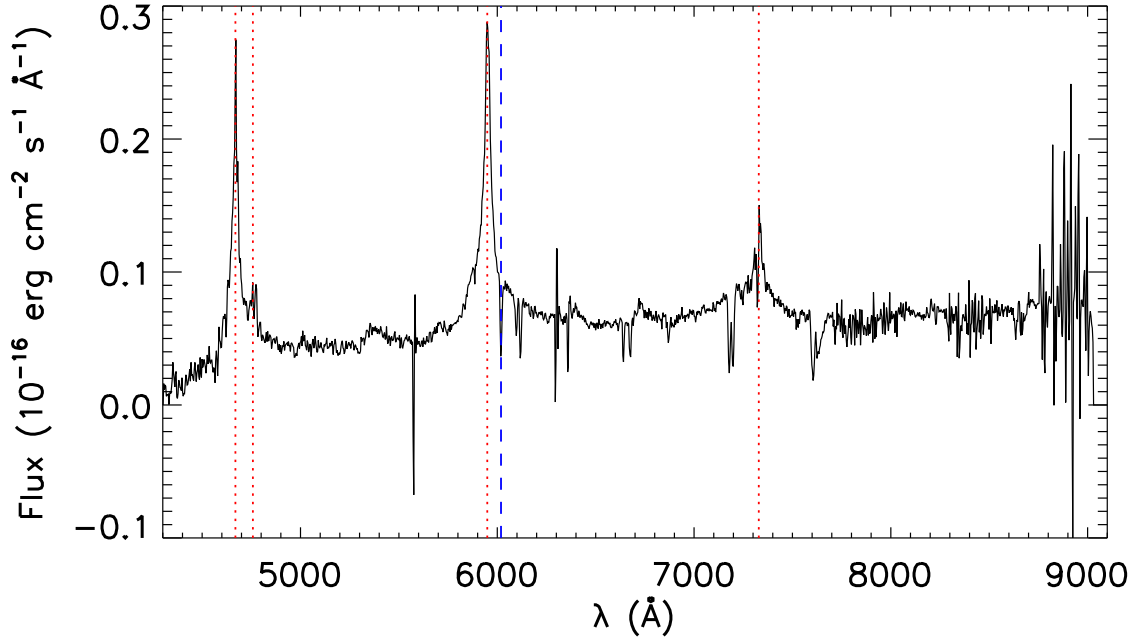}
\caption{VLT FORS2 long slit spectrum of the optical counterpart to 2XMM
J134404.1$-$271410.  We identify the broad emission lines (dotted red 
lines) as (from low to high wavelength) Lyman $\alpha$,
N {\sc v}, C {\sc iv} and C {\sc iii}.  From these we calculate a
redshift of $\sim 2.84$. 
Several absorption features can also be observed. The (unmarked) narrow features at $\sim 5579$ and 6302 {\angstrom}, and the broad feature around 7600 {\angstrom} are telluric. 
We tentatively identify the 6019 {\angstrom} absorption line (blue dashed line) as being consistent with neutral sodium in IC 4320, whilst the other unidentified absorption features may originate in one or more absorbers along the line-of-sight to the optical emission.
}
\label{spectrum}
\end{center}
\end{figure*}

The VLT FORS2 spectrum is shown in Figure \ref{spectrum}.  There are
clearly a number of broad emission lines, which suggest that the source
contains a type-1 active galactic nucleus (AGN).  We fitted these
emission lines with Gaussian profiles using {\sc
splat},\footnote{\url{http://star-www.dur.ac.uk/~pdraper/splat/splat-vo/
splat-vo.html}} and measured their wavelengths  and line widths. The
centroids of the emission lines are constrained to be at $4669 \pm 1$,
$4758 \pm 3$, $5949 \pm 1$ and $7329 \pm 4$ \angstrom, with FWHMs of $54
\pm 3$, $60 \pm 10$, $64 \pm 3$ and $90 \pm 10$ {\angstrom}
respectively.  These lines are consistent with the Lyman $\alpha$ 1216
\angstrom, N {\sc v} 1240 \angstrom, C {\sc iv} 1549 \angstrom~and C
{\sc iii} 1909 {\angstrom} lines respectively, if at a redshift of $\sim
2.84$, with velocity dispersions of $\sim 3000~{\rm
km~s^{-1}}$.  We can therefore confidently exclude 2XMM
J134404.1$-$271410 from being a HLX associated with the galaxy IC 4320;
instead we identify it as a background QSO.  We calculate a luminosity
distance of $\sim 2.4 \times 10^4~{\rm Mpc}$ to the QSO, for $H_{\rm 0}
= 71$, $\Omega_{\rm M} = 0.27$ and $\Omega_{\rm vac} = 0.73$.  At this
distance, 2XMM J134404.1$-$271410 has a peak observed 0.3--10 keV (rest
frame 1.15--38.4 keV) luminosity of $\sim 2.2 \times 10^{46}~{\rm
erg~s^{-1}}$ (calculated using the peak 0.3--10 keV observed flux from
\citealt{sutton_etal_2012}); this is not unusually luminous for a
redshift $\sim 3$ source (cf. \citealt{ueda_etal_2014}).

It is possible to obtain an estimate of the black hole mass from the FWHM of the C {\sc iv} line and the continuum luminosity. Here, we do this for 2XMM J134404.1$-$271410 using the relation from \cite{vestergaard_and_peterson_2006}: \begin{multline}
{\rm log}~M_{\rm BH}({\rm C \textsc{~iv}}) 
= 
\rm{log}\left\{
\left[
\rm{FWHM} (\rm{C \textsc{~iv}}) \over {1000~\rm{km~s^{-1}}} 
\right]^2 
\times\right.
\\
\left.
\left[ 
{\lambda L_{\lambda}(1450 \angstrom)} \over {10^{44}~\rm{erg~s^{-1}}}
\right]^{0.53}
\right\} 
+ (6.73 \pm 0.01).
\end{multline}
This process is slightly complicated by the fact that the 1450~{\angstrom} luminosity at the redshift of the emission lines is close to the 5579~{\angstrom} skyline. As such, we estimate the continuum luminosity at a rest frame wavelength of 1450~{\angstrom} by extrapolating the spectrum from either side of the 5579~{\angstrom} absorption feature. We estimate a black hole mass of $\sim 2 \times 10^8 M_{\odot}$, which is indicative of an Eddington ratio of $\sim 1$. However, we note that the C {\sc iv} black hole mass estimate is subject to a high degree of systematic uncertainty. Not only is there no common procedure for fitting the emission line, as it is confused with other sources of emission (e.g. \citealt{fine_etal_2010}), but it may also be subject to non-gravitational effects \citep{baskin_and_laor_2005}.

In addition to the emission lines, there are also a number of absorption features in the FORS2 spectrum. As above, we measured their wavelengths using {\sc splat}, and checked whether these features were consistent with originating at the emission line redshift, or that of IC 4320. After eliminating a number of telluric features, we tentatively identifed an absorption feature at $6019.2 \pm 0.4$ {\angstrom} as originating from neutral Na in IC 4320 (Figure \ref{spectrum}). We suggest that the other unidentified features may originate in one or more intervening absorbers.

\section{Discussion}\label{discussion}

\begin{figure*}
\begin{center}
\includegraphics[width=18cm]{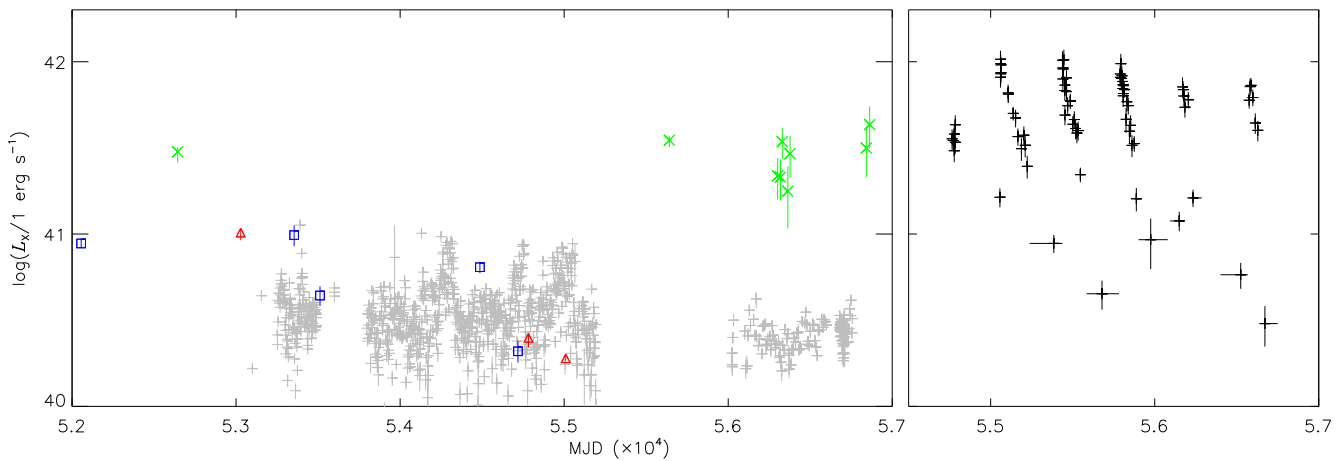}
\caption{({\it left}) Long term light curve showing 2XMM
J134404.1$-$271410 (green diagonal crosses) under the incorrect
assumption that it is at the distance of IC 4320, and three of the
remaining good HLX candidates.  The sources are Cartwheel N10 (blue
squares), 2XMM J011942.7$+$032421 in NGC 470 (red triangles) and M82 X-1
(grey crosses).  The data for 2XMM J134404.1$-$271410 are 0.3--10 keV
observed luminosities. They include the {\it XMM-Newton} and {\it
Chandra} detections reported in \protect\cite{sutton_etal_2012}, plus
several {\it Swift} XRT observations. 0.3--10 keV count rates were
extracted from the {\it Swift} data, then  converted to luminosities
using {\sc pimms} and an assumed absorbed power-law spectrum, with
parameters from \protect\cite{sutton_etal_2012}.  The data for 2XMM
J011942.7$+$032421 are 0.3--10 keV observed luminosities taken from
\protect\cite{sutton_etal_2012}.  The Cartwheel N10 data were estimated
from Figure 4 of \protect\cite{pizzolato_etal_2010}, and converted to
0.3--10 keV observed luminosities using a distance of 122 Mpc
\protect\citep{wolter_and_trinchieri_2004} and the power-law spectral
fits reported in \protect\cite{pizzolato_etal_2010}.  The M82 X-1 data
are from {\it RXTE} PCA and {\it Swift}  XRT observations taken in
photon counting mode.  For the {\it RXTE} observations, we used count
rates from the archival mission-long light curve.  We then estimated a
conversion factor to 0.3--10 keV observed luminosities using {\sc
pimms}, for an absorbed power law model with $N_{\rm H} = 1.12 \times
10^{22}~{\rm cm^{-2}}$ and $\Gamma = 1.67$
\protect\citep{kaaret_etal_2006}.  For the {\it Swift} data, we
extracted 2--10 keV background subtracted spectra using {\sc xselect},
and fitted these with an absorbed power-law model in {\sc xspec}.  The
absorption column density and power-law spectral index were fixed to the
values from \protect\cite{kaaret_etal_2006}.  The choice of the 2--10
keV bandpass was due to the necessity of reducing contamination from
diffuse emission in the M82.  We extrapolated the absorbed power-law to
the full 0.3--10 keV band, to estimate the observed flux.  M82 X-1 is
not resolved from other nearby point sources by {\it Swift} or {\it
RXTE\/}, so the values that we obtain are effectively upper-limits on
the flux from X-1.  The second brightest point source in the extraction
region is M82 X-2, which at its peak is around half as luminous as X-1
\protect\citep{miyawaki_etal_2009}. ({\it right}) Light curve showing
the FRED profile of ESO 243-49 HLX-1, shown on the same time scale. 
0.3--10 keV count rates were extracted from {\it Swift} XRT data, and
were grouped in to bins of 50 counts. These count rates were converted
to estimated luminosities by assuming a bimodal distribution of X-ray
spectra, with typical parameters taken from
\protect\cite{servillat_etal_2011}. Below $5 \times 10^{-3}~{\rm
cts~s^{-1}}$ an absorbed power-law spectrum was assumed, with $N_{\rm H}
= 3 \times 10^{20}~{\rm cm^{-2}}$ and $\Gamma = 2.1$.  At higher count
rates, ESO 243-49 HLX-1 was assumed to have an absorbed
multi-colour-disc spectrum, with $N_{\rm H} = 3 \times 10^{20}~{\rm
cm^{-2}}$ and $kT = 0.22~{\rm keV}$.}
\label{lightcurve}
\end{center}
\end{figure*}

Although now recognised as highly interesting objects, given that they
appear good candidates for hosting IMBHs, the number of {\it bona
fide\/} HLXs remains small.  With our identification of one of the best
remaining HLX candidates, 2XMM J134404.1--271410, as a background QSO,
their scarcity is emphasised further.  In this section we therefore
consider what the properties of the few remaining objects in this class
might tell us about their nature.

But first, it is interesting with hindsight to speculate whether 2XMM J134404.1$-$271410 
could have been rejected as a potential ULX based on its X-ray to optical flux ratio.
We use the optical magnitude in the V-band, along with the {\it Chandra} flux 
(calculated from a power-law model; see \citealt{sutton_etal_2012} for details) 
to estimate the X-ray to optical flux ratio of the target. We use the 
\cite{stocke_etal_1991} definition of the flux ratio (${\rm log} 
(f_{\rm X}/f_{\rm opt}) \equiv {\rm log}~f_{\rm 0.3-3.5 keV} + m_{V}/2.5 + 5.37$), 
to obtain ${\rm log} (f_{\rm X}/f_{\rm opt}) = 1.67 \pm 0.2$. This sits between 
the ranges of X-ray to optical flux ratios for typical 
AGN (-1 -- 1.2; \citealt{stocke_etal_1991}) and ULXs (2 -- 3; 
\citealt{tao_etal_2011}). 
Such a high flux ratio could be indicative
of an obscured QSO at high redshift, as X-ray absorption 
decreases strongly at high energies, whilst dust extinction increases toward the UV. Indeed, \cite{fiore_etal_2003} report that the X-ray to optical flux ratio roughly scales as $(1+z)^{3.6}$ (albeit in different X-ray and optical bands). If this relation approximately holds for the definition of the X-ray to optical flux ratio that we use here, then we would expect ${\rm log} (f_{\rm X}/f_{\rm opt}) \sim 0$ at $z \sim 0$, which is entirely consistent with a QSO.
However, we note that a few ULXs do have values of 
${\rm log} (L_{\rm X}/L_{\rm opt}) < 2$ \citep{tao_etal_2011, heida_etal_2013}, thus 
neither identification could be categorically ruled out based on X-ray to optical 
flux ratio alone.

Before its exclusion as a HLX, 2XMM J134404.1$-$271410 provided an
interesting bridge between ESO 243-49 HLX-1 and the other HLX
candidates.  Aside from being intermediate in peak luminosity between
ESO 243-49 HLX-1 and other HLXs, it was also the only other HLX
candidate to (apparently) reside in an early-type galaxy.  In fact, both
its apparent host (IC 4320) and that of ESO 243-49 HLX-1 are lenticular
galaxies that host a dust lane (e.g. \citealt{farrell_etal_2012}; cf.
Fig.~2).  In contrast, the other good, relatively nearby HLX candidates
are located in late-type hosts \citep{RC3}: a ring galaxy (Cartwheel
N10); an interacting SA(rs)b (NGC 470, 2XMM J011942.7$+$032421; 
with the association between the galaxy and HLX being confirmed by \citealt{gutierrez_and_moon_2014}); and an
edge-on I0 galaxy (M82 X-1).  In fact, the location of these three less
luminous HLXs is strongly reminiscent of the ordinary ULX population, as
is the case for most of the extreme ULXs ($L_{\rm X} > 5 \times 10^{40}
\rm ~erg~s^{-1}$; \citealt{sutton_etal_2012}), and indeed both the
Cartwheel and M82 also host several fainter ULXs (e.g.
\citealt{gao_etal_2003,wolter_and_trinchieri_2004,kaaret_etal_2006}). 
So, by excluding the IC 4320 object, ESO 243-49 HLX-1 appears more
strongly distinguished from the other HLXs in terms of its host system.

One characteristic that should perhaps have marked 2XMM
J134404.1$-$271410 out as different to other HLXs was its relatively
persistent X-ray luminosity, not varying much from $\sim 3 \times
10^{41} \rm ~erg~s^{-1}$ (under the misapprehension it is associated
with IC 4320; \citealt{sutton_etal_2012}; cf. Fig.~\ref{lightcurve}). 
This behaviour is not seen in the other reasonably well-studied HLXs,
that we show are all far more variable over comparable time scales in
Fig.~\ref{lightcurve}.   Of these, ESO 243-49 HLX-1 is unique in
displaying a well constrained FRED-like outburst cycle
\citep{lasota_etal_2011}. It is also the only remaining source that is
observed to exceed $\sim 2 \times 10^{41}~{\rm erg~s^{-1}}$
\citep{farrell_etal_2009}.  M82 X-1 has been extensively monitored with
both {\it RXTE\/} and {\it Swift\/}; it exhibits multiple epochs of
flaring (e.g. \citealt{pasham_and_strohmayer_2013}), which can peak at
$\sim 10^{41} \rm ~erg~s^{-1}$, and was reported to have a possible 62
day period in {\it RXTE} observations \citep{kaaret_and_feng_2007},
although it is now thought that this is probably super-orbital in nature
\citep{pasham_and_strohmayer_2013}. The X-ray lightcurves of Cartwheel
N10 and 2XMM J011942.7$+$032421 are not as well sampled.  However,
similarly to M82 X-1, we do know that they are not persistently
hyperluminous, rather they can appear with substantially diminished
fluxes in subsequent epochs \citep{wolter_etal_2006,
pizzolato_etal_2010, sutton_etal_2012}.

It is also important to consider the X-ray spectra of the HLX
candidates.  2XMM J134404.1$-$271410 was considered a good IMBH
candidate, as it had a low/hard state-like spectrum whilst appearing to
be a HLX (although in retrospect this was also consistent with an AGN-like spectrum).  
In contrast, 2XMM J011942.7$+$032421 has been reported to
have a curved, disc-like X-ray spectrum when it is at its most luminous,
but to appear more power-law-like when faded below its peak luminosities
\citep{sutton_etal_2012}.   The X-ray spectrum of Cartwheel N10 is
inconclusive; archival data are relatively low quality and statistically
well-fitted by both power-law and disc models
\citep{pizzolato_etal_2010}.  However, M82 X-1 does also show a trend of
disc-like hard curvature when it is most luminous
\citep{chiang_and_kong_2011, sazonov_etal_2014}, with several authors
arguing this is indicative of a sub-Eddington low/hard -- high/soft
state transition (e.g.
\citealt{feng_and_kaaret_2010,chiang_and_kong_2011}).  It is notable,
however, for both M82 X-1 and 2XMM J011942.7$+$032421 the disc-like
states have temperatures $\sim 1$~keV, more indicative of a stellar-mass
black hole than a large IMBH.\footnote{We note however that very high disc temperatures 
are seen in a few Galactic black hole binaries in the steep power-law state 
(e.g., $kT = 2.7$ -- 3.8 keV in 4U 1630-􏰀47; \citealt{tomsick_etal_2005}). This 
would scale with the black hole mass as $M^{-1/4}$, thus corresponds to a disc 
temperature of $\sim 1$ keV for a $\sim 10^3 ~M_{\odot}$ IMBH.}

It is interesting to speculate, in light of 2XMM J134404.1$-$271410 
being excluded from the HLX population, what the characteristics of the
remaining sources tells us about the physical nature of HLXs. In ESO
243-49 HLX-1 there is arguably strong evidence from both the
luminosity of the object and its behaviour for the presence of a
$10^4$--$10^5~M_{\odot}$ IMBH (e.g. \citealt{davis_etal_2011,
webb_etal_2012}; although see \citealt{lasota_etal_2015} for another
view on this object). The case for IMBHs in the other HLX candidates
is not always as strong, with in particular M82 X-1 having many
various black hole mass estimates (see
e.g. \citealt{miyawaki_etal_2009}; \citealt{pasham_etal_2014}).  The
exclusion of 2XMM J134404.1$-$271410 from the class of HLXs serves
primarily to highlight the gulf in peak luminosity and behavioral
properties between ESO 243-49 HLX-1 and the fainter HLXs.  We note
that the latter objects share a similar peak luminosity ($L_{\rm X}
\sim 10^{41}~{\rm erg~s^{-1}}$; Figure~\ref{lightcurve}), albeit in
two cases from very sparsely sampled lightcurves, which is very close
to the peak luminosity predicted for massive stellar black holes that
are maximally accreting (see Section~\ref{intro}).  They also share their
host environments - young, star-forming regions - with the bulk of the
less luminous ULX population.  Their spectral properties are also
interesting in this sense - in super-Eddington models, the spectra are
predicted to become more disc-like at the highest luminosities as the
radiatively-driven wind envelops the objects and Compton
downscattering in the outflow dominates \citep{kawashima_etal_2012};
this may be what is observed in M82 X-1 and 2XMM J011942.7+032421.
(Interestingly, this phenomenon may also already have been seen in
some standard ULXs as they reach their peak luminosities,
e.g. \citealt{vierdayanti_etal_2010, pintore_and_zampieri_2012,
walton_etal_2014}).  This dichotomy in observed characteristics
therefore leads us to speculate that we could be seeing two separate
populations amongst the small number of HLXs: ESO 243-49 HLX-1 stands
alone as an outstanding IMBH candidate, while the other objects may
represent the absolute luminosity peak of the `normal' X-ray binary
population, powered by maximal hyper-Eddington accretion onto the
largest stellar remnant black holes.

\section{Conclusions}

In conclusion, in this paper we have demonstrated that 2XMM
J134404.1$-$271410 is a background QSO and not a HLX associated with IC
4320.  This causes the already small number of known HLX candidates in
the local Universe to decline further, but perhaps more importantly it
serves to emphasise the gulf in properties between the prototypical IMBH
candidate, ESO 243-49 HLX-1, and the handful of other objects in the HLX
class.  We have argued that this may be due to a real physical
difference between HLX-1 and the other objects; but clearly this is
highly speculative, being based on a small number of rather poorly
observed objects.  Future missions with both the survey capability to
detect more candidate HLXs (e.g. {\it eRosita\/}) and the collecting
area and instruments to study this class in far greater detail (e.g.
{\it Athena\/}) are required to make real progress in confirming or
refuting our suspicions.

\section*{Acknowledgements}

ADS and TPR acknowledge funding from the Science and Technology
Facilities Council as part of the consolidated grants ST/K000861/1 and
ST/L00075X/1.  This work is based on observations made with ESO
telescopes at the La Silla Paranal Observatory under programme IDs
090.D-0300(A) and 092.D-0212(A).  It is also based in part on
observations made by the {\it Chandra} X-ray Observatory, and made use
of data supplied by the UK {\it Swift} Science Data Centre at the
University of Leicester.

\bibliography{refs}
\bibliographystyle{mn2e}

\bsp

\label{lastpage}

\end{document}